\documentclass[rnote]{aa}
\usepackage{graphicx}
\usepackage{natbib}
\usepackage{amssymb}
\bibpunct{(}{)}{;}{a}{}{,}

\begin{document}

\title{Detection of X-rays from the jet-driving symbiotic star \\ Hen\,3-1341}

\author{Matthias Stute\inst{1}
\and
Gerardo J. M. Luna\inst{2,3}
\and
Ignazio F. Pillitteri\inst{3}
\and
Jennifer L. Sokoloski\inst{4}
}

\offprints{Matthias Stute, \\ 
\email{matthias.stute@tat.physik.uni-tuebingen.de}}

\institute{
Institute for Astronomy and Astrophysics, Section Computational
Physics, Eberhard Karls Universit\"at T\"ubingen, Auf der Morgenstelle 10, 
72076 T\"ubingen, Germany
\and
Instituto de Astronom\'ia y F\'isica del Espacio (CONICET-UBA), 
Casilla de Correo 67, Suc. 28 (C1428ZAA), Ciudad Aut\'onoma de 
Buenos Aires, Argentina
\and
Harvard-Smithsonian Center for Astrophysics, 60 Garden St.,
MS 15, Cambridge, MA, 02138, USA
\and
Columbia Astrophysics Laboratory, 550 W. 220th Street, 1027
Pupin Hall, Columbia University, New York, NY 10027, USA
}

\date{Received 09 February 2012 ; accepted 10 May 2013}

\abstract
{Hen\,3-1341 is a symbiotic binary system consisting of a white dwarf and a red 
giant star that is one of about ten symbiotics that show hints of jets. The 
bipolar jets have been detected through displaced components of emission lines 
during its outburst from 1998 to 2004. These components disappeared when 
Hen\,3-1341 reached quiescence. On February 23, 2012, Hen\,3-1341 started a new 
outburst with the emergence of new bipolar jets on March 3, 2012.}
{We observed Hen\,3-1341 during quiescence with {{\em XMM-Newton}} in March 
2010 with an effective exposure time of 46.8 ks and with Swift on March 
8--11, 2012 as ToO observations with an effective exposure time of 10 ks in 
order to probe the interaction of the jet with the ambient medium and also the 
accretion onto the white dwarf.}
{We fitted the {{\em XMM-Newton}} X-ray spectra with XSPEC and examined the 
X-ray and UV light curves.}
{We report the detection of X-ray emission during quiescence from Hen\,3-1341 
with {{\em XMM-Newton}}. The spectrum can be fitted with an absorbed 
one-temperature plasma or an absorbed blackbody. We did not detect Hen\,3-1341 
during our short Swift exposure. Neither periodic or aperiodic X-ray 
nor UV variability were found.} 
{Our {{\em XMM-Newton}} data suggest that interaction of the residual jet with 
the interstellar medium might survive for a long time after outbursts and might 
be responsible for the observed X-ray emission during quiescence. Additional 
data are strongly needed to confirm these suggestions.}


\maketitle

\section{Introduction}

Hen\,3-1341 ($=$ V2523\,Oph $=$ SS73\,75) was discovered by \citet{Hen76}, and 
its classification as a symbiotic star followed the spectroscopic observations 
by \citet{All78}. The high-excitation conditions found by Allen were later 
confirmed by \citet{GMC97}, whose optical and ultraviolet spectra show strong 
emissions in the Balmer continuum, by the N V, [Fe VII], He II lines, and the 
symbiotic band at 6830 \AA\ due to Raman scattering by neutral hydrogen. 
\citet{MuS99} derived a spectral type of M4 for the cool giant. Its optical 
colors are typical of symbiotic stars harboring a very hot and luminous white 
dwarf and infrared colors appropriate to a cool giant without circumstellar 
dust. In quiescence, Hen\,3-1341 resembles Z\,And, the prototype of symbiotic 
stars \citep{Ken86}. 

About 200 symbiotic stars are known \citep[e.g.][]{BMM00}, but jets have been 
detected at different wavelengths only in about ten of them \citep{BSK04}. 
While Hen\,3-1341 was in outburst, \citet{TMM00} discovered a jet with a radial 
velocity of 820 km s$^{-1}$ in high-resolution spectra leading to emission 
components displaced on both sides of the main emission lines. \citet{MSH05} 
followed the evolution of the jet emission components and reports their 
disappearance when the system returned to quiescence. On February 23, 2012, 
Hen\,3-1341 started a new outburst with the emergence of new bipolar 
jets on March 3, 2012 \citep{MMS12, MVD12}.

X-ray observations are an excellent probe of the bow and internal shocks of 
the jet that emit soft X-rays (photon energies $\lesssim 2$ keV) and also the 
central parts of the jet engine, where gas is being accreted to power the jet, 
leading to hard (photon energies $\gtrsim 2$ keV) and/or soft X-ray emission. 
Up to now, only in R\,Aqr \citep{KPL01,KAK07} and CH\,Cyg \citep{GaS04, KCR07,
KGC10} have jets from symbiotic stars been resolved in X-rays. All known jet 
sources, when observed in X-rays, show soft components with $k\,T\lesssim2$ 
keV: R\,Aqr \citep{KPL01,KAK07}, CH\,Cyg \citep{GaS04, KCR07,KGC10}, MWC\,560 
\citep{StS09}, V1329\,Cyg \citep{SLS11}, RS\,Oph \citep{LMS09}, AG\,Dra 
\citep{GVI08} and Z\,And \citep{SKE06}. Furthermore, the three objects CH\,Cyg, 
R\,Aqr, and MWC\,560 also emit hard X-rays \citep{MIK07,NDK07,StS09}. Z\,And 
showed hard emission in only one of three observations \citep{SKE06}.

The paper is organized as follows. In Section \ref{sec_obs}, we show details of 
the observations and the analysis of the data. After that we describe the 
results in Section \ref{sec_res}. We end with a discussion and conclusions in 
Section \ref{sec_dis}. 

\section{Observation and analysis} \label{sec_obs}

\subsection{{\em XMM-Newton} observations during quiescence}

We observed the field of Hen\,3-1341 for $\sim$58 ks with {\em XMM-Newton} in 
March 2010 using the EPIC instrument operating in full window mode and with the 
{\em Medium} thickness filter. Simultaneously, we used the Optical Monitor (OM).
All the data reduction was performed using the Science Analysis Software 
(\textsc{SAS}) software package\footnote{http://xmm.vilspa.esa.es/} version 
10.0. We removed events collected during high background intervals. We further 
filtered events to keep only those with {FLAG=0} and PATTERN in 0--4 (only 
single and double events) for pn and 0--12 for MOS, as prescribed in the SAS 
guide. The resulting exposure time after these steps is 47 ks with pn and 58 ks 
with MOS. 

After first inspection of the data, only a weak source was present at the 
position of Hen\,3-1341, therefore we used a source detection algorithm based 
on a multiscale wavelet convolution \citep{DMM97a,DMM97b} specifically tailored 
for EPIC {\em XMM-Newton} cameras. In particular it allows the detection using 
the weighted sum of the images obtained with pn and MOS CCDs. We set the 
threshold for source detection at $4.5\sigma$ of local background to retain, on 
a statistical basis, at most one spurious source per field.

The source spectra and light curves were accumulated from a circular region of 
12'' radius (240 pixels, Fig. \ref{fig_image}) centered on Hen\,3-1341 using 
the coordinates from \textsc{SIMBAD}. We extracted the background spectra and 
light curves from a source-free circular region with a radius of 30'' (600 
pixels) on the same chip, at the same distance from the readout node as the 
region from which we extracted the source counts. Spectral redistribution 
matrices and ancillary response files were generated using the \textsc{SAS} 
scripts \texttt{rmfgen} and \texttt{arfgen}, and spectra were fed into the 
spectral fitting package 
\textsc{XSPEC}\footnote{http://heasarc.gsfc.nasa.gov/docs/xanadu/xspec/} 
v12.6.0. Due to the low number of counts, we did not group the 
counts and analyzed the ungrouped spectrum using the Cash fit statistics. For 
timing analysis, photon arrival times were converted to the solar system 
barycenter using the SAS task \texttt{barycen}. 

\subsection{Swift observations during outburst}

We observed the field of Hen\,3-1341 for $\sim$10 ks with Swift in March 
2012 as a triggered ToO observation after the reported emergence of new bipolar 
jets about two weeks earlier. The XRT observed in the PC (photon counting) mode.
To search for X-ray emission from Hen\,3-1341, we first concatenated 
the event files from three visits using the \texttt{evselect} tool, built 
and combined individual exposure maps using the \texttt{xrtexpomap} and 
\texttt{ximage} tools. Finally, we built a concatenated image and then used 
the \texttt{ximage} tool to search for source emission. 

We also obtained Swift/UVOT images from which we extracted fluxes in the UVM2, 
UVW2, and UVW1 filters using a 5" source extraction region, while background 
flux was extracted from an annulus of the 10" and 20" inner and outer radii, 
respectively, around the source.

\section{Results} \label{sec_res}

\subsection{{\em XMM-Newton}}

\subsubsection{Images}

\begin{figure}[!htb]
  \centering
  \includegraphics[width=\columnwidth]{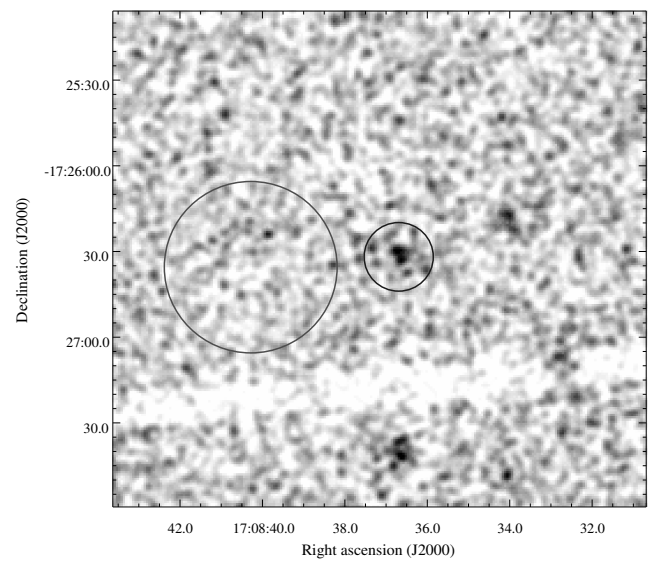}
  \caption{{\em XMM-Newton} EPIC MOS1 + MOS2 + pn image in $0.3-8.0$ keV 
    centered on Hen\,3-1341. Extraction regions for the source and background 
    events are overplotted.}
  \label{fig_image}
\end{figure}

{\em XMM-Newton} detected the source with a significance of 6.2 $\sigma$ in all 
EPIC images (Fig. \ref{fig_image}). With the pn camera, we detected 564 counts 
in the background region, so we expect about 90 background counts within the 
135 counts in the source region. Therefore the source has been detected with 4.7
$\sigma$ above the background. In the MOS 1 camera, we find 141 background and 
38 source counts, i.e. a detection with 3.2 $\sigma$. In the MOS 2 camera, we 
find 125 background and 32 source counts, i.e. a detection with 2.66 $\sigma$. 
The detected position agrees perfectly with the SIMBAD position of Hen\,3-1341 
and that of the closest source in the 2MASS All Sky catalog of point sources 
\citep{SCR06}. Other sources are considerably shifted with respect to our 
detected source. Thus it is very likely that the detected X-rays are indeed 
emitted by Hen\,3-1341. The average magnitudes in the optical filters U, UVW1, 
and UVM2 are 12.8, 12.9, and 13.9, respectively. 

\subsubsection{Spectra}

Because of the low number of counts, we grouped the counts only for 
plotting purposes and not for fitting the spectrum. Furthermore, we used the 
Cash fit statistic as implemented in XSPEC, which allows for the 
analysis of background subtracted Poisson-distributed data, as discussed in 
Appendix B of the XSPEC manual.

\begin{figure}[!htb]
  \centering
  \includegraphics[width=\columnwidth]{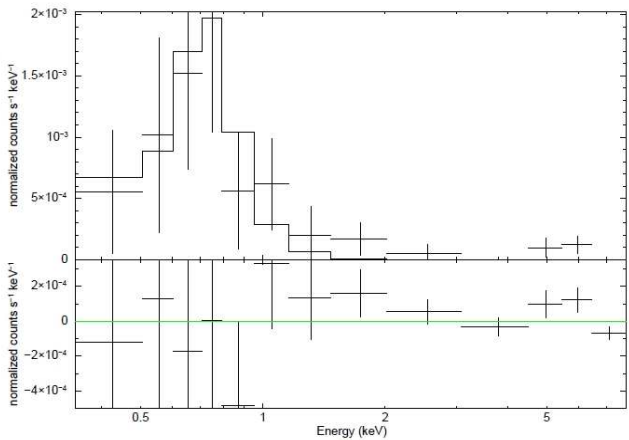}
  \caption{Observed pn spectrum of Hen\,3-1341, together with a model 
    consisting of an absorbed one-temperature plasma (\texttt{wabs(apec)}).}
  \label{fig_spectralfit1}
\end{figure}

The X-ray spectra of Hen\,3-1341 can be described with an absorbed 
single-temperature plasma (\texttt{wabs(apec)} in \textsc{XSPEC}). We find 
$N_H \gtrsim 8\times10^{21}$ cm$^{-2}$ and $k\,T = 0.32^{+0.33}_{-0.12}$ keV 
(Fig. \ref{fig_spectralfit1}). The total absorbed flux between 0.3--10 keV is 
$1.66\times10^{-15}$ ergs cm$^{-2}$ $s^{-1}$. 

We also fitted the spectrum with an absorbed blackbody model 
(\texttt{wabs(bbody)}). The fit also describes the spectra well (reduced 
$\chi^2 = 0.482$, 5 d.o.f.), and we find $N_H = (5.04^{+89.8}_{-5.11})\times10^{20}$ 
cm$^{-2}$ and $k\,T = 0.24^{+0.19}_{-0.19}$ keV. This corresponds to a temperature 
of 2.8 MK with an error range of $T = (0.6-5)$ MK, which is better 
constrained than in the first model. The total absorbed flux between 0.15--10 
keV is $(1.84\pm0.03)\times10^{-15}$ ergs cm$^{-2}$ s$^{-1}$.

The slope of the ultraviolet spectral energy distribution clearly 
shows that the optical and ultraviolet fluxes are more consistent
with nebular emission than blackbody emission from a hot white dwarf.

\subsubsection{Light curves}

To study the X-ray variability, we used the Kolmogorov-Smirnov 
\citep[KS, e.g. ][]{PTV93} and Kuiper tests \citep{Kui60} and compared the 
cumulative distributions of a constant model, source, and background events. The
KS and Kuiper test did not detect variability (at 3$\sigma$ level) in the event 
arrival rate from any of the EPIC cameras. 

We also examined the OM photometry with exposure times of 1500s and 3000s. The
measured rms variations in the count rates are consistent with those expected 
from Poisson statistics.

\subsection{Swift}

We did not detect Hen\,3-1341 with our short-exposure Swift observation.
The 3$\sigma$ upper limit is $1.3\times 10^{-3}$ counts/sec. 
Again using the two best-fit models mentioned above, this count rate would 
correspond to an absorbed flux of about $2.8\times10^{-14}$ ergs cm$^{-2}$ 
s$^{-1}$ and an unabsorbed flux of about $5\times10^{-14}$ ergs cm$^{-2}$ 
s$^{-1}$.

With UVOT, we find average UV count rates of $74\pm8$ counts/sec (UVM2, 2246 
\AA), $224\pm9$ counts/sec (UVW1, 2600 \AA), and $141\pm8$ (UVW2, 1928 \AA). 
Converted to magnitudes, these values correspond to $12.14\pm0.03$, 
$11.56\pm0.03$, and $12.01\pm0.03$, respectively. 

\section{Discussion and conclusions} \label{sec_dis}

Hen\,3-1341 shows a soft X-ray emission. One possible explanation
can be given in light of simulations of jets in symbiotic stars 
\citep{Stu06,StS07} or protostellar jets \citep[e.g.][]{BOM10,BOM11} where soft 
X-ray emission arises from internal shocks and the bow shock. The velocity of 
the shock, $v_{\rm shock}$, can be derived using the measured temperature of the 
component and the following relation (assuming strong shock conditions):
\begin{eqnarray}
T_{\rm post\,shock} &=& \frac{3}{16}\,\frac{\mu\,m_P}{k_B}\,v_{\rm shock}^2 
\nonumber \\
&=& 0.105\,\textrm{keV}\,\left(\frac{v_{\rm shock}}{\textrm{300 km s$^{-1}$}}
\right)^2 
\end{eqnarray}
with $m_p$ the proton mass, $k_B$ the Boltzmann constant, and $\mu=0.6$ the 
mean particle weight. Therefore the observed soft component temperature 
range using the APEC model corresponds to a shock velocity range of
400--750 km s$^{-1}$. \citet{TMM00} discovered a jet with a 
radial velocity of 820 km s$^{-1}$ while Hen\,3-1341 was in outburst. The 
displaced spectral component then gradually disappeared when the system 
returned to quiescence in 2004 \citep{MSH05}. Since then the outflow has likely 
decelerated by the interstellar medium. Thus it is possible that the soft 
component is emitted by residuals of this interaction.

Although we know that a jet had been present in this object, other scenarios 
for explaining the emission of soft X-rays are possible:
\begin{itemize}
\item One scenario is photospheric emission from a hot white dwarf. The observed
UV flux could in principle be produced by a white dwarf with a radius of 0.14 
R$_\odot$ and a luminosity of 3000 L$_\odot$, as proposed by \citet{MSH05}, 
but the X-ray emission cannot. An effective temperature of 2.8 MK, which 
would be required to produce the observed X-ray emission, is inconsistent with 
the optical emission-line spectrum \citep{MSH05}. 
\item Another possible source of the detected X-rays may be colliding winds 
\citep{MWJ97}, since soft X-ray emission has also been detected in objects 
without jets \citep{MWJ97}. We disfavor this picture because no wind signatures 
have been detected in spectroscopic observations during quiescence or 
outburst \citep{WDS93, TMM00} -- only signatures of a bipolar jet during the
outbursts in 1999 and 2012 \citep{TMM00,MVD12}. 
\end{itemize} 

We performed hydrodynamical simulations of the propagating jet with parameters 
representative of Hen\,3-1341 (Stute et al., in prep.), in which we varied the 
jet mass-loss rate following the observed AAVSO light curve. Temperature and 
density maps from our simulations are used, together with emissivities 
calculated with the atomic database ATOMDB, for estimating the X-ray luminosity 
emitted by the jet. Although we made simple assumptions, we can reproduce 
the order of magnitude of the observed X-ray luminosity.

The X-ray and the optical light curves are consistent with a constant flux. 
The measured rms variations in the count rates are consistent with those 
expected from Poisson statistics.

Using the effective exposure time of 46.8 ks in all instruments, the 
sensitivity of {\em XMM-Newton} (Watson et al. 2001) gives an upper limit of 
about 10$^{-14}$ erg s$^{-1}$ cm$^{-2}$ in the hard band ($\gtrsim 2$ keV). Using 
a distance of 3.1 kpc \citep{GMC97}, the flux upper limit gives an upper limit 
for the the hard X-ray luminosity of about $10^{31}$ erg s$^{-1}$. \citet{MSH05} 
estimats a white dwarf mass of 0.4 M$_\odot$, a radius of 0.1 R$_\odot$  and 
an accretion rate of $5\times10^{-8}$ M$_\odot$ yr$^{-1}$, which would lead to 
\begin{equation}
L_x \lesssim \frac{G\,M_{WD}\,\dot M}{2\,R_{WD}} = 1.2\times10^{34} \,
\textrm{erg s$^{-1}$}\,.
\end{equation}
Therefore either the accretion rate has decreased substantially or the accretion
luminosity from the boundary layer is emitted in other wavelength bands besides 
the X-rays; i.e., the accretion-disk boundary layer is optically thick, as in 
Mira -- another jet-producing symbiotic star with very low X-ray flux 
\citep{SoB10}. In any case, the accretion rate estimate agrees with 
Hen\,3-1341 being in quiescence.

Unfortunately, we did not detect Hen\,3-1341 with our short-exposure 
Swift observation. The derived upper limit shows that the soft X-ray 
component has not increased its flux by a factor higher than about 20. This
neither supports nor rejects our conclusions. The Swift observations 
took place about one week after the reported emergence of the jets. In our 
simulations we found that it takes the X-ray luminosity up to 20 days to 
increase significantly after a new emerged jet has started interacting with the 
surrounding medium \citep[][Stute et al., in prep.]{StS07}. Furthermore, the 
magnitude of the increase depends on the jet parameters.

We have to note that uncertainties in the distance might change the flux 
levels; however, \citet{GMC97} estimate an error of their derived distance of 
only 4\%.

New data are strongly needed to confirm our conclusion.

\acknowledgements
We acknowledge helpful comments and suggestions by an anonymous referee. 
This work is based on observations obtained with {\em XMM-Newton}, an ESA 
science mission with instruments and contributions directly funded by ESA Member
States and the USA (NASA). We thank NASA for funding this work through 
{\em XMM-Newton} AO-8 awards NNX09AP88G to GJML and JLS, and NNX10AK31G to JLS. 
We gratefully acknowledge the effort of the entire Swift team for the execution 
of our ToO request. This publication makes use of data products from the Two 
Micron All Sky Survey, which is a joint project of the University of 
Massachusetts and the Infrared Processing and Analysis Center/California 
Institute of Technology, funded by the National Aeronautics and Space 
Administration and the National Science Foundation. We acknowledge with thanks 
the variable star observations from the AAVSO International Database 
contributed by observers worldwide and used in this research.

\end{document}